\documentclass[final,3p,times,twocolumn]{elsarticle}
 \biboptions{comma,sort&compress}
\usepackage{here}
 \usepackage{graphicx}

\usepackage{amssymb}
\usepackage{amsthm}
\usepackage{amsmath}

\def\nin{\noindent}
\def\beq{\begin{equation}}
\def\eeq{\end{equation}}
\def\bea{\begin{eqnarray}}
\def\eea{\end{eqnarray}}

\journal{Nuc. Phys. (Proc. Suppl.)}

\begin{document}

\begin{frontmatter}



\title{Transverse Spin Physics at COMPASS}

 \author{Christian Schill \\ 
 on behalf of the COMPASS collaboration}
  \address{Physikalisches Institut der
Albert-Ludwigs-Universit\"at Freiburg\\
Hermann-Herder-Str. 3, 79104 Freiburg, Germany}
\ead{Christian.Schill@cern.ch}



\begin{abstract} 
\noindent 

The investigation of transverse
spin and transverse momentum effects in deep inelastic scattering is one of the
key physics programs of the COMPASS collaboration. 

 Three channels have been
analyzed at COMPASS to access the transversity distribution function: The
azimuthal distribution of single hadrons, involving the Collins fragmentation
function, the azimuthal dependence of the plane containing hadron pairs,
involving the two-hadron interference fragmentation function, and the
measurement of the transverse polarization of $\Lambda$ hyperons in the final
state.  

Azimuthal asymmetries in unpolarized semi-inclusive deep-inelastic
scattering give important information on the inner structure of the nucleon as
well, and can be used to estimate both the quark transverse momentum $k_T$ in
an unpolarized nucleon and to access the so-far unmeasured Boer-Mulders
function. COMPASS has measured these asymmetries using spin-averaged $^6LiD$
data. 

\end{abstract}


\end{frontmatter}


\section{Introduction}
\nin

Most of our knowledge of the inner structure of the nucleon is encoded in
parton distribution functions. They are used to describe hard scattering
processes involving nucleons. While a lot of understanding has been achieved on
the longitudinal structure of a fast moving nucleon, very little is known about
its transverse structure \cite{Anselmino0}. Recent data on single spin
asymmetries in semi-inclusive deep-inelastic scattering (SIDIS) off
transversely polarized nucleon targets \cite{COMPASS,HERMES} triggered a lot of
interest towards the transverse momentum dependent and spin dependent
distribution and fragmentation functions \cite{Bacchetta}. 

The SIDIS cross-section in the one-photon exchange approximation contains eight
transverse-momentum dependent distribution functions \cite{Bacchetta3}. Some of
these  can be extracted in SIDIS measuring the azimuthal distribution of the
hadrons in the final state \cite{Aram}. Three distribution functions  survive
upon integration over the transverse momenta: These are the quark momentum
distribution $q(x)$, the helicity distribution $\Delta q(x)$, and the
transversity distribution $\Delta_T q(x)$ \cite{Collins}. The latter is defined
as the difference in the number density of quarks with momentum fraction $x$
with their transverse spin parallel to the nucleon spin and their transverse
spin anti-parallel to the nucleon spin \cite{Artru}. 

To access transversity in SIDIS, one has to measure the quark polarization,
i.e. use the so-called 'quark polarimetry'. Several techniques are used at
COMPASS:  a measurement of the single-spin asymmetries (SSA) in the azimuthal
distribution of the final state hadrons (the Collins asymmetry), a measurement
of the SSA in the azimuthal distribution of the plane containing  final state
hadron pairs (the two-hadron asymmetry), and a measurement of the polarization
of final state hyperons (the $\Lambda$-polarimetry). In these proceedings, I
will focus on new results for the two-hadron asymmetry, while results for the
other channels are shown elsewhere \cite{COMPASS2}.

The chiral-odd  Boer-Mulders function is of special interest among the other
transverse-momentum dependent distribution functions \cite{Boer}. It describes
the transverse parton polarization in an unpolarized hadron. The Boer-Mulders
function generates azimuthal asymmetries in unpolarized SIDIS, together with
the so-called Cahn effect \cite{Cahn}, which arises from the fact that the
kinematics is non-collinear when $k_T$ is taken into account.

\section{The COMPASS experiment}
\nin

COMPASS is a fixed target experiment at the CERN SPS accelerator with a wide
physics program focused on the nucleon spin structure and on hadron
spectroscopy. COMPASS investigates transversity and the transverse momentum
structure of the nucleon in semi-inclusive deep-inelastic scattering. A
$160$~GeV muon beam is scattered off a transversely polarized $NH_3$ or
$^6LiD$  target. 

The scattered muon and the produced hadrons are detected in a
wide-acceptance two-stage spectrometer with excellent particle identification
capabilities \cite{Experiment}. The data  with a
transversely polarized $NH_3$ target shown here were taken in the $2007$ run.

\section{Two-hadron asymmetry}
\nin

The chiral-odd transversity distribution $\Delta_T q(x)$ can be measured
in combination with the chiral-odd polarized two-hadron interference fragmentation 
function $H^{\sphericalangle}_1 (z,M^2_{inv})$ in SIDIS. $M_{inv}$ is the invariant mass of the
$h^+h^-$ pair. 
The fragmentation of a transversely polarized quark into two unpolarized
hadrons leads to an azimuthal modulation in $\Phi_{RS} = \phi_R + \phi_s -
\pi$ in the SIDIS cross section. 
Here $\phi_R$ is the azimuthal angle between $\vec R_T$ and the lepton scattering plane and 
$\vec R_T$ is the transverse component of $\vec R$ defined as:
\begin{equation}
\vec R = (z_2\cdot \vec p_1 - z_1 \cdot \vec p_2)/(z_1+z_2).
\end{equation}
 $\vec p_1$ and $\vec p_2$ are the momenta in the laboratory frame of $h^+$
and $h^-$ respectively. This definition of $\vec R_T$ is invariant
under boosts along the virtual photon direction.

The number of produced oppositely charged hadron pairs $N_{h^+h^-}$ can be written as:
\begin{equation}
N_{h^+h^-} =N_0 \cdot ( 1 + f \cdot P_t \cdot D_{NN} \cdot A_{RS} \cdot \sin \Phi_{RS} \cdot \sin
\theta).
\end{equation}
Here, $\theta$ is the angle between the momentum vector of $h^+$ in
the center of mass frame of the $h^+h^-$-pair and the momentum vector of
the two hadron system \cite{Bacchetta}. 

The measured amplitude $A_{RS}$ is proportional to the product of the
transversity distribution and the polarized two-hadron interference fragmentation function 
\begin{equation}
A_{RS} \propto \frac {\sum_q e_q^2 \cdot \Delta_T q(x) \cdot H^{\sphericalangle}_1(z,M^2_{inv})}
 {\sum_q e_q^2 \cdot q(x) \cdot D_q^{2h}(z,M^2_{inv})}.
\end{equation}
$D_q^{2h}(z,M^2_{inv})$ is the unpolarized two-hadron interference fragmentation function.
The polarized two-hadron interference fragmentation function
$H^{\sphericalangle}_1$ can be expanded in the relative
partial waves of the hadron pair system, which up to the
p-wave level gives~\cite{Bacchetta}:
\begin{equation}
H^{\sphericalangle}_1 = H^{\sphericalangle,sp}_1 + \cos  \theta \cdot H^{\sphericalangle,pp}_1.
\end{equation}
Where $H^{\sphericalangle,sp}_1$ is given by
the interference of $s$ and $p$ waves, whereas the function
$H^{\sphericalangle,pp}_1$ originates from the interference of two $p$
waves with different polarization. For this analysis the results are
obtained by integrating over $\theta$. The $\sin \theta$
distribution  is strongly peaked at one
and the $\cos \theta$ distribution is symmetric around zero.

Both the interference fragmentation function
$H_1^\sphericalangle(z,M_{inv}^2)$ and the corresponding spin averaged fragmentation
function into two hadrons $D_q^{2h}(z, M_{inv}^2)$ are unknown, and need to be measured in $e^+e^-$
annihilation or to be evaluated using models \cite{Bacchetta,Jaffe,Bianconi,Radici}.

\section{Results for hadron pairs}
\nin

\begin{figure}\hspace*{-0.5cm}
     \includegraphics[width=1.1\columnwidth]{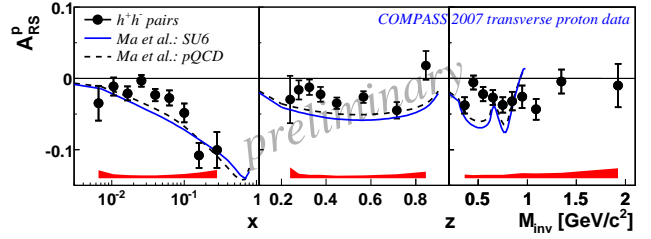}
     \caption{Two-hadron asymmetry $A_{RS}$ as a function of $x$, $z$ and $M_{inv}$, compared to 
     predictions of \cite{Ma}. 
     The lower bands indicate the
     systematic uncertainty of the measurement.}
	\label{pic:results}
\end{figure}

The two-hadron asymmetry as a function of $x$, $z$ and $M_{inv}$ is shown in
Fig.~\ref{pic:results}. A strong asymmetry in the valence $x$-region is observed,
which implies a non-zero transversity distribution and a non-zero polarized two
hadron interference fragmentation function  $H^{\sphericalangle}_1$. In the invariant
mass binning one observes a strong signal around the $\rho^0$-mass and the asymmetry
is negative over the whole mass range. 

The lines are calculations from Ma {\it et
al.}, based on a SU6 and a pQCD model for transversity \cite{Ma}. The
calculations can describe the magnitude and the $x$-dependence of the measured
asymmetry, while there are discrepancies in the $M_{inv}$-behavior.

\newpage

\section{Azimuthal asymmetries in DIS off an unpolarized target}
\nin

The cross-section for hadron production in lepton-nucleon DIS $\ell N
\rightarrow \ell' h X$ for unpolarized targets and an unpolarized or
longitudinally polarized beam has the following form~\cite{bacchetta2}:

\begin{equation}
\begin{array}{lcr}\displaystyle
\frac{d\sigma}{dx dy dz d\phi_h dp^2_{h,T}} =
  \frac{\alpha^2}{xyQ^2}
\frac{1+(1-y)^2}{2} \cdot\\[2ex] \displaystyle [ F_{UU,T} + 
  F_{UU,L} + \varepsilon_1 \cos \phi_h F^{\cos \phi_h}_{UU} \\[2ex]
   + \varepsilon_2 \cos(2\phi_h) F^{\cos\; 2\phi_h}_{UU}
   + \lambda_\mu
  \varepsilon_3
  \sin \phi_h F^{\sin \phi_h}_{LU} ]
\end{array}
\end{equation}
where $\alpha$ is the fine structure constant. 
$F_{UU,T}$,  $F_{UU,L}$, $F^{\cos \phi_h}_{UU}$,  $F^{\cos\;
  2\phi_h}_{UU}$ and $F^{\sin \phi_h}_{LU}$ are structure functions. Their 
first and second subscripts indicate the beam and target polarization,
respectively, and the last subscript denotes, if present, the
polarization of the virtual photon.  $\lambda_\mu$ is the 
longitudinal beam polarization and: 
\begin{equation}
\begin{array}{rcl}
\varepsilon_1 & = & \displaystyle\frac{2(2-y)\sqrt{1-y}}{1+(1-y)^2} \\[2ex]
\varepsilon_2 & = & \displaystyle\frac{2(1-y)}{1+(1-y)^2} \\[2ex]
\varepsilon_3 & = & \displaystyle\frac{2 y \sqrt{1-y}}{1+(1-y)^2}
\end{array}
\end{equation}
are depolarization factors. The Boer-Mulders parton distribution function
contributes to 
the $\cos \phi_h$ and the $\cos 2\phi_h$ moments as well, together with the
 Cahn effect~\cite{Cahn} which arises from the fact that the kinematics is
non collinear when
the $k_\perp$ is taken into account, and with the
perturbative gluon radiation, resulting in order $\alpha_s$ QCD processes. pQCD
effects become important for high transverse momenta $p_T$ of the produced
hadrons.

\section{Analysis of unpolarized asymmetries}
\nin

To obtain an unpolarized data sample, data taken with a longitudinally polarized
or a transversely polarized $^6LiD$ target in the year 2004 have both been spin-averaged. 

In the measurement of unpolarized asymmetries a Monte Carlo simulation is used
to correct for acceptance effects of the detector. The SIDIS event generation
is performed by  the LEPTO generator~\cite{lepto}, the experimental setup and the
particle interactions in the  detectors are simulated by the COMPASS Montecarlo
simulation program COMGEANT. 

The  acceptance of the detector as a function of the azimuthal angle $A(\phi_h)$ is then calculated as the
ratio of  reconstructed over generated events for each bin of $x$, $z$ and $p_T$ in which the
asymmetries are measured. The measured distribution, corrected for acceptance, is fitted with the
following functional form:
\[
\begin{array}{lll}
N(\phi_h) &=&N_0 \left( 1 + A^D_{\cos \phi}  \cos \phi_h + \right.  \\&& \left. A^D_{\cos 2\phi}
\cos 2\phi_h
 +  A^D_{\sin \phi} \sin \phi_h \right) 
\end{array}
\]
The contribution of the acceptance corrections to the systematic error was 
studied in detail.

\section{Results for unpolarized asymmetries}
\nin

\begin{figure}
\begin{center}\hspace*{-0.5cm}
\includegraphics[width=1.1\columnwidth]{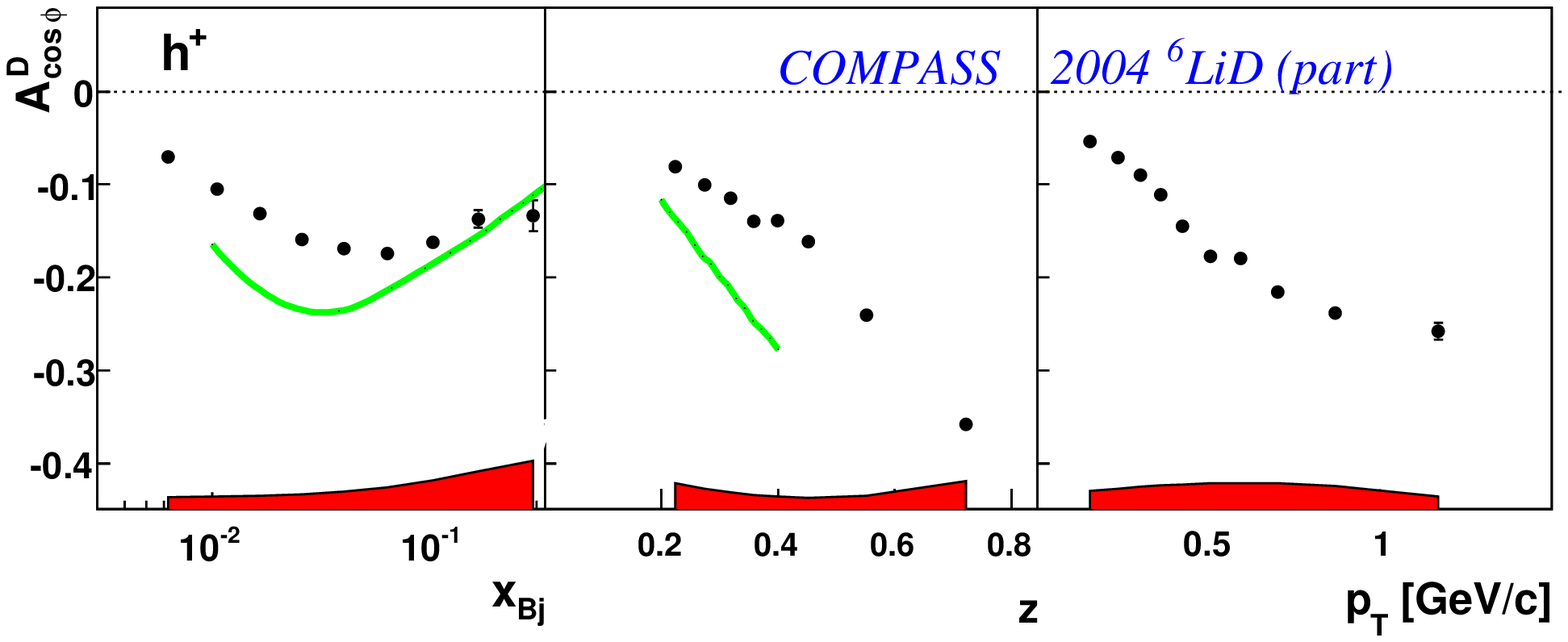}
\hspace*{-0.5cm}
\includegraphics[width=1.1\columnwidth]{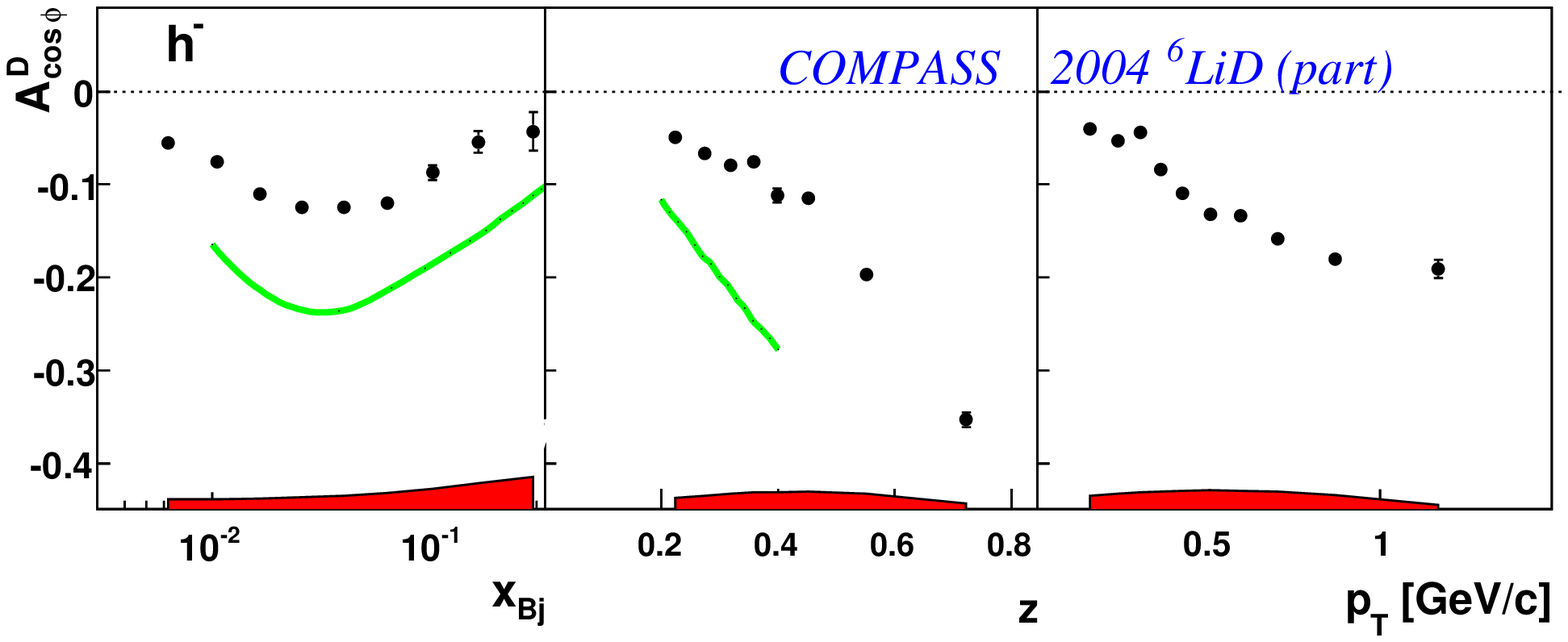}
\end{center}
\caption{$\cos \phi_h$ asymmetries from COMPASS deuteron data
for positive (upper row) and negative (lower
raw) hadrons; the asymmetries are divided by the kinematic factor
$\varepsilon_1$ and the bands indicate the size of the systematic uncertainty. 
The superimposed curves are the values predicted by~\protect\cite{anselmino2}
taking into account the Cahn effect only.
}
\label{f:cosphi}
\end{figure}

The $\sin \phi_h$ asymmetries measured by COMPASS, not shown here,  are compatible
with zero, at the present level of statistical and systematic errors, over the
full range of $x$, $z$ and $p_T$ covered by the data.

The $\cos \phi_h$ asymmetries extracted from COMPASS deuteron data
are shown in Fig.~\ref{f:cosphi} for positive (upper row) and negative (lower
row) hadrons, as a function of $x$, $z$ and $p_T$. The bands indicate the size
of the systematic error. The asymmetries show the same trend for positive and
negative hadrons with  slightly larger absolute values for  positive hadrons. 
Values as large as 30$-$40\% are reached in the last point of the $z$ range. 
The theoretical
prediction~\cite{anselmino2} in Fig.~\ref{f:cosphi} takes into account the Cahn
effect only, which
does not depend on the hadron charge. The Boer-Mulders parton distribution function 
is not
considered in this prediction. 

\begin{figure}
\hspace*{-0.5cm}
\includegraphics[width=1.06\columnwidth]{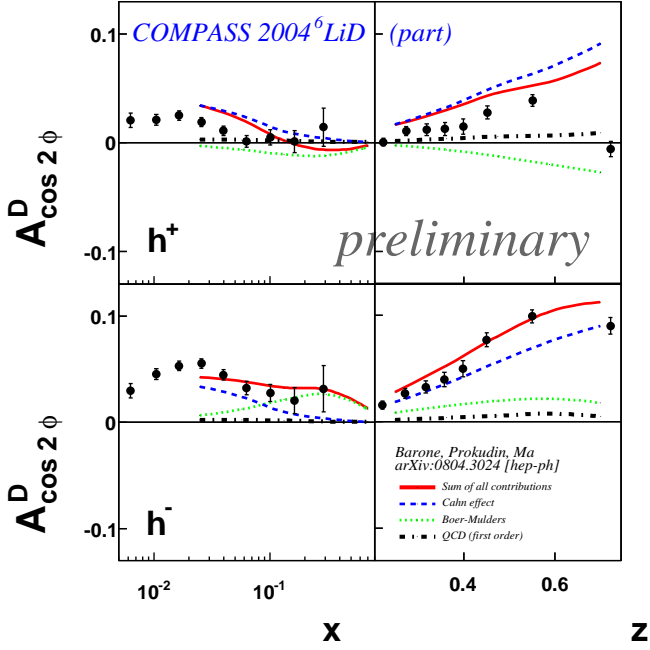}
\caption{$\cos 2 \phi_h$ asymmetries from COMPASS deuteron data
for positive (upper row) and negative (lower
raw) hadrons; the asymmetries are divided by the kinematic factor
$\varepsilon_1$ and the bands indicate size of the systematic error. 
}
\label{f:cos2phi}
\end{figure}

The $\cos 2 \phi_h$ asymmetries are shown in Fig.~\ref{f:cos2phi} together with the
theoretical predictions of~\cite{barone}, which take into account the kinematic
contribution given by the Cahn effect, first order pQCD (which, as expected, is
negligible in the low $p_T$ region), and  the Boer-Mulders parton distribution
function (coupled to the Collins fragmentation function), which gives a different
contribution to positive and negative  hadrons. 

In~\cite{barone}, the Boer-Mulders
parton distribution function is assumed to be proportional to the Sivers function as
extracted from preliminary HERMES data. The COMPASS data show an  amplitude
different  for positive and negative hadrons, a trend which confirms the theoretical
predictions. There is a satisfactory agreement between the data points and the model
calculations, which hints to a non-zero Boer-Mulders parton distribution function.

\section{Summary and Outlook}
\nin

New preliminary results for the two-hadron azimuthal asymmetry at COMPASS in
semi-inclusive deep-inelastic scattering off a transversely polarized proton target
have been presented. For $x>0.05$, an asymmetry different from zero and
increasing with increasing $x$-Bjorken has been observed. 

The measured unpolarized azimuthal asymmetries on a deuteron target show large
$\cos\phi_h$ and $\cos 2\phi_h$ moments which can be qualitatively described in
model calculations taking into account the Cahn effect and the intrinsic $k_T$
of the quarks in the nucleon and the Boer-Mulders structure function.

With a full-year transverse-target running in $2010$, COMPASS will significantly
increase its statistical precision in all measurements of transverse-spin
dependent asymmetries.

\section*{Acknowledgments}
\nin
This work has been supported by the German BMBF.












\end{document}